# Optically enhanced second harmonic generation in silicon oxynitride thin films via local layer heating

Jakub Lukeš,[1,2] Vít Kanclíř,[1,2] Jan Václavík,[1] Radek Melich,[1] Ulrike Fuchs,[3] Karel Žídek[1,*]

**AFFILIATIONS**

[1]*Research Center TOPTEC, Institute of Plasma Physics of the Czech Academy of Sciences, Za Slovankou 1782/3, 182 00 Prague, Czech Republic*

[2]*Technical University in Liberec, Faculty of Mechatronics, Informatics and Interdisc. Studies, Studentská 1402/2, 461 17 Liberec, Czech Republic*

[3] *Asphericon GmbH, Stockholmer Str. 9, 07747 Jena, Germany*

*Corresponding author: zidek@ipp.cas.cz

## Abstract

Strong second harmonic generation (SHG) in silicon nitride has been extensively studied—among others, in terms of laser-induced SHG enhancement in $Si_3N_4$ waveguides. This enhancement has been ascribed to the all-optical poling induced by the coherent photogalvanic effect. Yet, an analogous process for $Si_3N_4$ thin films has not been reported. Our article reports on the observation of laser-induced 3-fold SHG enhancement in $Si_3N_4$ thin films. The observed enhancement has many features similar to all-optical poling, such as highly nonlinear power dependence, cumulative effect, or connection to the $Si_3N_4$-Si interface. However, identical experiments for low-oxygen silicon oxynitride thin films lead to complex behavior, including laser-induced SHG reduction. By a thorough experimental study, the observed results were ascribed to heat-induced SHG variation caused by multiphoton absorption. Such behavior indicates that the origin of optically-induced SHG enhancement in $SiO_xN_y$-Si structures can be a complex interplay of various phenomena.

# Introduction

Silicon nitride ($Si_3N_4$), as well as silicon oxynitrides ($SiO_xN_y$), have attracted attention from many prospective applications in optics. These materials are used for optical coating as a means of creating layers with a graded refractive index.[1] Nevertheless, recently the research of silicon nitride has been motivated predominantly by its nonlinear optical properties, including strong second harmonic generation (SHG) with high potential in integrated photonics.[2]

In recent years, many studies have reported on the strong laser-induced enhancement of SHG in silicon nitride waveguides[3-5] and microresonators.[6,7] This enhancement has been ascribed to the effect of all-optical poling, where the driving mechanism was the so-called coherent photogalvanic effect. This effect induces internal local electric fields in the material and, therefore, allows efficient frequency doubling in centrosymmetric material via third-order nonlinearity (EFISH).[3,4] The photogalvanic effect occurs when a sample is exposed to the fundamental laser beam and its second harmonic, which can either originate from an external source or be generated in the sample itself.

The optically-induced SHG variation has also been reported for oxidized Si surfaces. The variation has been ascribed to the multiphoton electron and hole injection across the $Si$-$SiO_2$ interface. However, this time-dependent SHG has been restricted only to oxide layers below 10 nm and disappeared for thicker layers.[8]

For $SiO_xN_y$ optical thin films exceeding 10 nm in thickness, SHG efficiency has been accepted to be determined by the deposition process and thin film structuring. For instance, SHG intensity has been shown to be promoted by a stoichiometry of $Si_3N_4$,[9] a targeted deposition of $Si_3N_4$ and $SiO_xN_y$ structures with enhanced residual stress,[10] or accumulated fixed charges on layer interfaces.[10] In contrast to the waveguides and microresonators, optically induced SHG enhancement in optical thin films has not been previously reported.

In this article, we report on our observations of optically-induced SHG variation in silicon nitride and oxynitride thin films on a silicon substrate. In particular, we observed a strong 3-fold SHG enhancement on $Si_3N_4$ layers. Some characteristics of the enhancement closely resemble the coherent photogalvanic effect, including the highly nonlinear power dependence or cumulative character of the SHG enhancement.[3,4] Our measurements also revealed that the SHG variation is not linked to any notable change in the layer refractive index or chemical composition.

However, as our study extended to silicon oxynitride layers, we noted a more complex behavior depending on the layer stoichiometry. In some cases, the illumination by IR femtosecond pulses even reduced the SHG intensity. The observed behavior was incompatible with the coherent photogalvanic effect.

Therefore, we carried out a set of experiments testing the presence of heat-induced changes in the sample, such as the effect of irradiation laser repetition rate or ex-situ sample annealing. The acquired experimental results fully confirmed our hypothesis. They indicate that the all-optical SHG enhancement in silicon nitride and oxynitride can be a complex process originating from an interplay of multiple phenomena.

# Results

### SHG variation in $SiO_xN_y$ thin films

As the first step, we studied laser-induced SHG enhancement in $Si_3N_4$ thin films deposited via dual ion beam sputtering on the Si substrate.[11] Throughout the article, we used p-polarized IR pulses at 1028 nm both for the sample irradiation and SHG measurement; the incident angle of the IR beam was 70 deg and we measured p-polarized SHG – see Methods for details. [12]

Laser-induced SHG variation measurements were implemented by placing a sample on an XY stage, which allowed us to irradiate the sample point-by-point and also to map the SHG intensity, as well as IR reflectivity. The laser-induced SH intensity variation was measured by illuminating several rectangular segments of the sample, using a different IR laser intensity for each segment—see Fig. 1(A). Subsequently, we performed one overall XY scan at a low laser intensity, which was used to record the SHG and reflected IR intensity— see Fig. 1(A).

We observed significantly enhanced SHG in the regions previously illuminated by high-intensity IR pulses—see Fig. 1A. The dependence of SHG enhancement on illumination power in Fig. 1B followed the $I^6$, which had been previously reported also for optical poling via the coherent photogalvanic effect.[3,4] The SHG enhancement was homogeneous across the illuminated area, which implies that it was not connected to any local contamination of the sample.

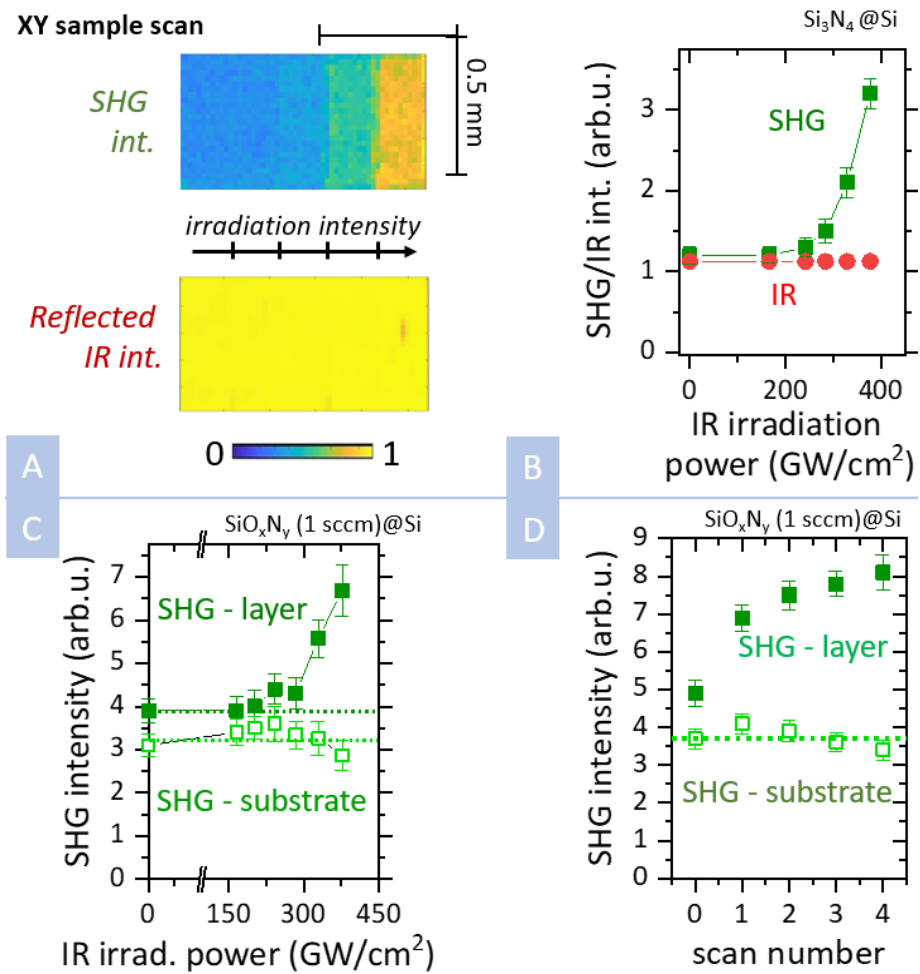

**Fig. 1**. Laser-induced SHG enhancement (A) XY spatial scan of a sample—measured SHG and reflected IR intensities on $Si_3N_4$ layer previously illuminated in segments with a laser intensity increasing from left to right. (B) Mean SHG (green squares) and reflected IR (red circles) intensity after the previous illumination for a set of irradiation intensities (peak powers). Data measured for $Si_3N_4$ layer (0 sccm, Si substrate, 1500 nm thickness). The "zero" irradiation intensity corresponds to an unexposed layer. (C) SHG enhancement measured on $SiO_xN_y$ layer (1 sccm, Si substrate, 1200 nm)—solid squares. Compared to SHG intensity on the adjacent Si substrate— open squares. (D) SHG enhancement dependence on the number of illumination scans over the same segment. Measured on $SiO_xN_y$ layer (1 sccm, Si substrate, 1200 nm)—solid squares. Compared to SHG intensity on the adjacent Si substrate— open squares.

In parallel to the SHG measurement, we also measured the IR reflectivity of the illuminated area (see Fig. 1A-B), which remained constant within the relative statistical error of 0.5%. Since we measured the p-polarization reflectance in the proximity of the Brewster angle, we can infer that there cannot be any major change in the refractive index of the layer or substrate. Analogous SHG enhancement was also observed for oxynitride thin films deposited on a silicon substrate with 1 sccm oxygen flow— see Fig. 1C. We denote silicon oxynitride layers by the flow of oxygen used during the layer deposition. While $\varphi(O_2)$ = 0 sccm corresponds to the pure $Si_3N_4$, the flow of $\varphi(O_2)$ = 3 sccm leads to the formation of nearly $SiO_2$ layers.[13] The estimate of stoichiometric factors for each sample is provided in Supplementary information, Section 1.

We also studied the case when a segment of a sample is irradiated multiple times— see Fig. 1D. We noticed that the effect is cumulative, i.e., SHG enhancement increases with a higher number of scans but it shows signs of saturation after a few repetitions.

Interestingly, we observed the laser-induced SHG variation only for the layers on the Si substrate, while the same layers deposited on the BK7 substrate did not show any sign of enhancement—see Fig. 2A. We confirmed this behavior for a variety of $SiO_xN_y$ thin films deposited within the same batch, which differed only by their substrate. Therefore, in the following text, we restricted ourselves to layers deposited on the Si substrate.

To exclude the possibility that the laser-induced changes occurred in the substrate only, we carried out measurements close to the layer edge, where we irradiated the layer and the bare substrate within a single experiment. The Si substrate without a layer showed only minor SHG changes within the statistical error of the measurement—see Fig. 1C–D. We also verified that the amplitude of the IR laser electric field was comparable for the bare substrate and the substrate below the layer. Therefore, the absence of SHG enhancement on the bare substrate shows that the SHG variation originated in the layer, and it requires the presence of the Si-layer interface.

The observed properties of SHG enhancement presented in Figs. 1 and 2A closely resemble the coherent photogalvanic effect reported previously on $Si_3N_4$ waveguides. The similarities include the $I^6$ nonlinear behavior, cumulative SHG enhancement, SHG enhancement without apparent change in the linear optical response of the layer and connection to the $Si$-$SiO_xN_y$ interface.

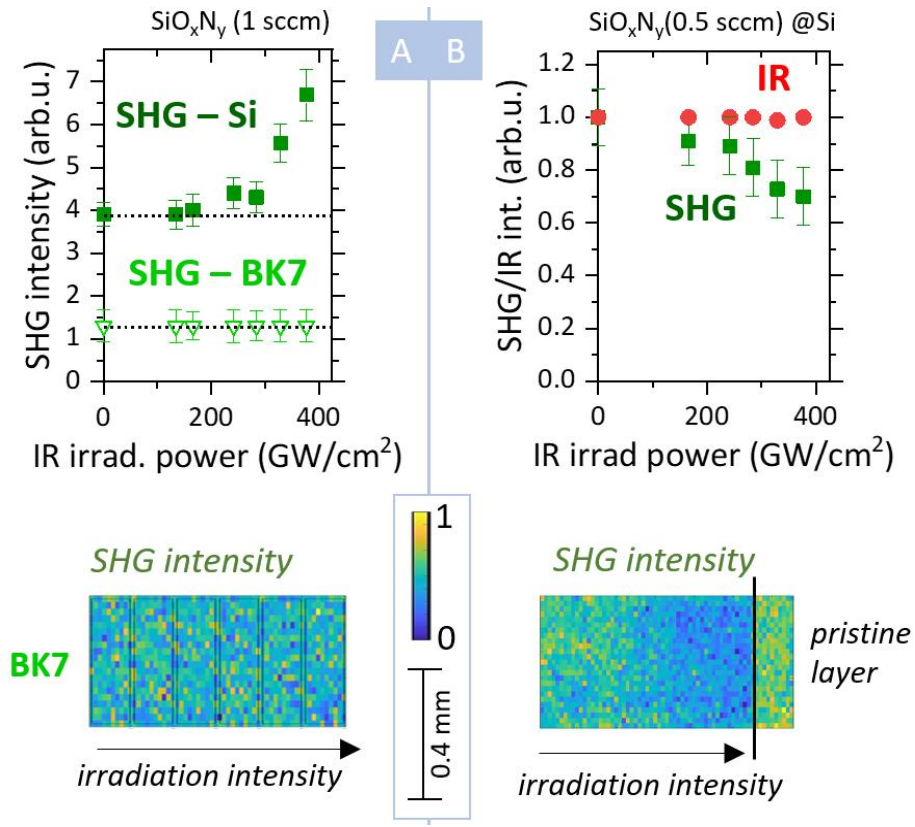

**Fig. 2**. (A) Comparison between SHG enhancement of SiO$_x$N$_y$ thin layer (1 sccm, 1200 nm) deposited on Si substrate (solid squares) and same layer deposited on BK7 substrate (open triangles). Lower part: SHG intensity map—XY scan of the illuminated layer on BK7. (B) SHG intensity (green squares) and reflected IR intensity (red circles) after laser illumination on SiO$_x$N$_y$ layer (0.5 sccm, 1200 nm, Si substrate)— dependence on the illumination power. Lower part: SHG intensity map —XY scan of the illuminated layer. The pristine (reference) area was not illuminated prior to the measurement.

The photoinduced SHG variation became more complex for the low-oxygen oxynitride thin film samples. We observed that for SiO$_x$N$_y$ layers with an oxygen flow of 0.25–0.5 sccm, the SH intensity decreased with the increasing IR illumination intensity—see Fig. 2B. This behavior was persistent for different deposited samples.

The SHG reduction on the low-oxygen oxynitride thin films in Fig. 2B contradicted the interpretation via optical poling. The coherent photogalvanic effect is inherently phase-matched and always leads to SHG enhancement.[3] Moreover, the previously reported methods of erasing the built-in internal electric field by illuminating the sample with an externally generated SH light have been proven not to affect the observed changes in our samples.[3]

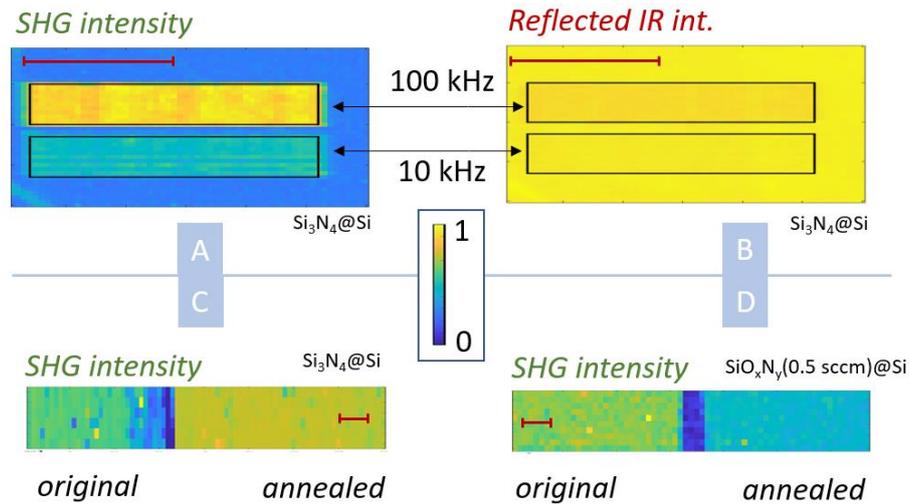

**Fig. 3.** (A) XY scan of SHG intensity of $Si_3N_4$ layer (0 sccm, 1500 nm, Si substrate) irradiated by two laser repetition rates: 100 kHz for 2 s (upper rectangular area) and 10 kHz for 20 s (bottom rectangular area), irradiation laser intensity 230 GW/cm$^2$. (B) XY scan of reflected IR intensity—same conditions and sample as in panel A. (C) Thermal enhancement of SHG for $Si_3N_4$ layer via ex-situ annealing: XY scan over two segments of the same sample—LHS: without heat treatment; RHS: 400°C annealed. (D) Thermal enhancement of SHG for $SiO_xN_y$ with sccm 0.5 via ex-situ annealing (LHS without heat treatment, RHS 400°C annealed). The red scale bars correspond to 0.5 mm for all panels.

## Heat-induced SHG variation

We propose a different phenomenon consistent with the presented experimental data as the source of the SHG variation – heat-induced changes triggered by a highly nonlinear absorption. To evaluate the possibility of a heat-induced sample transformation, we carried out a simple calculation, which provided us with a gross estimate of the temperature change induced by a single pulse—see Supplementary information, Section 4 for details. Depending on the absorption depth, we can estimate that a single pulse can increase the local temperature by tens of Kelvins at most but cannot by itself provide enough heat for layer restructuring. This implies that the potential heat-induced effect must be cumulative.

Therefore, we carried out sets of measurements where we varied the laser repetition rate. If a coherent photogalvanic effect or an analogous nonlinear process is responsible for the SHG enhancement, we should observe the same SHG enhancement when we keep a constant irradiation peak power and the same number of incident pulses on the spot. This would hold irrespective of the laser repetition rate. On the contrary, the heat-induced changes in a sample should strongly depend on the laser repetition rate, even if the total number of incident pulses is the same. This is because the long delay between the pulses allows for better heat dissipation between the pulses.

By carrying out the SHG enhancement measurement for 100 kHz and 10 kHz repetition rates where we maintained the same number of incident pulses, we could see that the enhancement is highly dependent on the laser repetition rate—see Fig. 3A. This result indicates that we observe heat-induced changes in the Si-SiO$_x$N$_y$ interface.

To confirm this conclusion, we conducted an experiment with ex-situ sample annealing. We cut a Si substrate with a Si$_3$N$_4$ layer into two parts. One of those parts was uniformly heated to 400°C for 30 minutes and then left to cool down at room temperature. Both parts of the sample were then placed into the setup and scanned as a single measurement. The results can be seen in Fig. 3C, where the annealed part of the sample is shown on the right-hand side, while the left-hand side is the part without thermal treatment. The average measured intensity of the SH of the annealed part is approximately 20% higher compared to the non-annealed part. At the same time, changes in reflected IR intensity are subtle.

Moreover, we conducted the same experiment on the SiO$_x$N$_y$ sample (0.5 sccm) for which we had previously observed a decrease in SHG due to irradiation with a laser beam - see Fig. 2B. In line with our expectations, here, the annealing had led to the reduction of SHG average intensity on the annealed part by approximately 30% in comparison to the part which had not been heat treated— see Fig. 3D.

## Discussion

In light of our results, we can safely ascribe the observed laser-induced SHG enhancement/reduction to localized layer heating. Even though the laser-induced heating probably takes place on the substrate-layer interface, heat conduction leads to an even temperature level across the whole layer depth, which we have verified by a simple finite-element method simulation. Due to the highly nonlinear character of the SHG enhancement ($I^6$ dependence), we can speculate that what we observed was the effect of multiphoton absorption since tunneling ionization typically becomes dominant only for light intensities several orders of magnitude higher.[14]

We can now turn to the identification of the mechanism behind the SHG enhancement itself, where we propose two viable scenarios of SHG variation: (i) formation of a new heat-induced sublayer in the sample, (ii) heat-induced restructuring of the layer bulk. To gain more insight into these options, we created an optical model of the sample—see Supplementary information, Section 5. The optical model studies the variation in the IR reflection of the sample induced by a change in the refractive index of the thin films. From the model, we can conclude that the observed subtle changes in IR reflectance put constraints on the two above-mentioned scenarios. The change in the refractive index can be very high (>0.1) only for the formation of a very thin layer (< 10 nm). For the layer bulk modification (> 50 nm thick), the refractive index is expected not to vary by more than 0.01.

The heat can, in principle, induce the formation of a very thin sublayer at the layer-substrate interface via atom diffusion across the interface. However, the diffusion coefficients of Si and N in $Si_3N_4$ and the corresponding diffusion length during the short illumination period of 2 s would reach a notable effect only when the local temperature in the layer was significantly above 1700°C,[15] i.e., exceeding the melting point of Si and approaching the melting point of $Si_3N_4$. This temperature level is not viable since we have observed that the layer can resist a significantly higher irradiation level and the related temperature without being destroyed. Diffusion of atoms (e.g., oxygen) across the layer-air interface cannot be responsible for the enhancement, as this process would be active for both Si and BK7 substrates.

Therefore, we propose the heat-induced restructuring of the layer bulk as the most viable explanation. We can speculate that the mechanical stress in the layer might be responsible for the SHG variation, as the stress: (i) highly affects the SHG efficiency due to symmetry breaking;[10] (ii) leads to low refractive index change;[16] (iii) its hysteresis is highly different for $Si_3N_4$ and $SiO_xN_y$ layers, as the character of the stress changes from compressive to tensile.[17] Our experiments with layer annealing at 400°C, which is sufficient to change the stress in the layer,[18] led to SHG modification in line with laser-induced changes. Nevertheless, other heat-induced mechanisms cannot be excluded.

In summary, we carried out a thorough investigation of light-induced SHG modification, which we observed, unlike other reports, on thin-film samples. While the nonlinear behavior and other aspects might suggest that we witnessed the coherent photogalvanic effect, we carried out a set of measurements that safely assigned the SHG modification to localized heating of the layer. The heating is causing the SHG to vary due to the restructuring of the layer bulk.

Our results provide a better understanding of the mechanism of SHG in $Si_3N_4$ and $SiO_xN_y$ thin-film layers on a Si substrate. We see that the SHG is facilitated primarily by changes within the layer volume, which can arise, for instance, due to symmetry breaking in the layer caused by mechanical stress or local inhomogeneities. This is in line with previous results of our group and Ning et al.,[9,12] who reported a strong SHG from thin films with a bulk dipolar character. While such a mode of SHG is not allowed in centrosymmetric media, the stress or the layer inhomogeneity lifts these constraints.

In general, for studying $Si_3N_4$ layers, it is highly beneficial to use a reduced laser repetition rate as a simple means of minimizing laser-induced layer modification.[12] Measurements of SHG under various laser repetition rates also present a simple way to verify the presence of heat-induced modification.

## Methods

### Second harmonic generation setup

The SHG setup used in this study was described in detail in Ref. 12. Amplified Yb:YAG fs pulses (4 µJ/pulse) at 1028 nm were directed into the SHG setup, where their intensity was modulated to the desired level. The IR pulses were focused on a sample (20 µm spot diameter) to generate SHG in the reflective geometry. Throughout the article, we used p-polarization IR light, incident angle 70 deg, and we detected p-polarization SHG radiation. Using a series of dichroic optics and color filters, we simultaneously detected the intensity of the reflected IR light and the generated SH radiation. Unless stated otherwise, the experiments were carried out at the laser repetition rate of 100 kHz.

### Sample preparation

Sample preparation was carried out using the dual ion beam sputtering described in detail in Ref. 12. A beam of $Ar^+$ ions (beam current 108 mA, beam voltage 600 V) sputtered Si atoms from a target onto substrates (Silicon or BK7), where the deposited atoms interacted with nitrogen and oxygen ions generated by an assistance ion beam (emission current 0.6 A, discharge voltage 70 V). By changing the ratio between the oxygen and nitrogen ion flux, we were able to vary the stoichiometry of the layers from the pure $Si_3N_4$ ($\varphi(O_2)$ = 0 sccm) through $SiO_xN_y$ to nearly $SiO_2$ layers ($\varphi(O_2)$ = 3 sccm). The estimate of stoichiometric factors is provided in Supplementary information (Section 1). We carried out a detailed study of the linear optical properties of the layers in Ref. 13. The thickness of the layers varied between 300 and 3500 nm.

### Laser-induced SHG variation

Laser-induced SHG variation measurements were implemented by placing a sample on an XY stage, which allowed us to scan the sample and irradiate it point by point. The laser-induced SH intensity variation was measured by illuminating several rectangular segments of the sample, using a different IR laser intensity for each segment—see Fig. 1(A).

During the irradiation, the segment was scanned spot-by-spot by the IR laser (1028 nm), irradiating each spot for 2 s unless stated otherwise. The irradiation intensity ranged from 170 to 430 $GW/cm^2$, corresponding to 35–100 $mJ/cm^2$/pulse. Subsequently, we performed one overall XY scan recording the SH and reflected IR intensity— see Fig. 1(A). The overall XY scan covered all previously irradiated segments and an adjacent reference area without prior irradiation. During the overall scan, the intensity of the IR beam was kept at a constant low level (170 $GW/cm^2$, 35 $mJ/cm^2$/pulse) to avoid additional sample changes. The data gained from the overall scan were then evaluated independently for each segment by calculating the average intensity and the standard deviation as an error estimation— see Fig. 1(B).

## Supplemental document

See the Supplementary Information document for supporting measurements and models.

## Acknowledgments

We gratefully acknowledge financial support of Ministerstvo školství, mládeže a tělovýchovy (Project No. CZ.02.1.01/0.0/0.0/16_026/0008390). We also thank Hana Libenská and Gleb Pokatilov for their help with sample characterization and heat conduction simulations.

## Ethics declarations

**Competing interest**

The authors declare no conflicts of interest.

## Data availability

Data underlying the results presented in this paper are not publicly available but may be obtained from the authors upon reasonable request.

## References

[1] A. M. Kaddouri, A. Kouzou, A. Hafaifa, and A. Khadir, "Optimization of anti-reflective coatings using a graded index based on silicon oxynitride," J. Comput. Electron. 18, 971 (2019).

[2] D. J. Blumenthal, R. Heideman, D. Geuzebroek, A. Leinse, and C. Roeloffzen, "Silicon nitride in silicon photonics," Proc. IEEE 106, 2209 (2018).

[3] O. Yakar, E. Nitiss, J. Hu and C.-S. Brès, "Coherent Photogalvanic Effect for Second-Order Nonlinear Photonics," arXiv:2203.06980v1 (2022)

[4] E. M. Dianov and D. S. Starodubov, "Photoinduced generation of the second harmonic in centrosymmetric media," Quantum Electron. 25, 395 (1995).

[5] D. D. Hickstein, D. R. Carlson, H. Mundoor, J. B. Khurgin, K. Srinivasan, D. Westly, A. Kowligy, I. I. Smalyukh, S. A. Diddams, and S. B. Papp, "Self-organized nonlinear gratings for ultrafast nanophotonics," Nat. Photonics 13, 494 (2019).

[6] X. Lu, G. Moille, A. Rao, D. A. Westly, and K. Srinivasan, "Efficient photoinduced second-harmonic generation in silicon nitride photonics," Nat. Photonics 15, 131 (2021).

[7] E. Nitiss, J. Hu, A. Stroganov, and C.-S. Brès, "Optically reconfigurable quasi-phase-matching in silicon nitride microresonators," Nat. Photon. 16, 134 (2022).

[8] J. Bloch, J. G. Mihaychuk, and H. M. van Driel, "Electron Photoinjection from Silicon to Ultrathin Si Films via Ambient Oxygen," Phys. Rev. Lett. 77, 920 (1996).

[9] T. Ning, H. Pietarinen, O. Hyvärinen, J. Simonen, G. Genty, and M. Kauranen, "Strong second-harmonic generation in silicon nitride films," Appl. Phys. Lett. 100, 161902 (2012).

[10] R. Wehrspohn, C. Schriever, and J. Schilling, "Inhomogeneous strain in silicon photonics," ECS Transactions 61, (2014).

[11] K. Žídek, J. Hlubuček, P. Horodyská, J. Budasz, and J. Václavík, "Analysis of sub-bandgap losses in $TiO_2$ coating deposited via single and dual ion beam deposition," Thin Solid Films 626, 60 (2017).

[12] N. K. Das, V. Kanclíř, P. Mokrý, and K. Žídek, "Bulk and interface second harmonic generation in the $Si_3N_4$ thin films deposited via ion beam sputtering," J. Opt. 23, 024003 (2021).

[13] V. Kanclíř, J. Václavík, and K. Žídek, "Precision of silicon oxynitride refractive-index profile retrieval using optical characterization," Acta Phys. Pol. A 140, 215 (2021).

[14] M. Ams, D. J. Little, and M. J. Withford, Laser Growth and Processing of Photonic Devices (Woodhead Publishing, 2012), Chap. 10.

[15] H. Schmidt, U. Geckle, and M. Bruns, "Simultaneous diffusion of Si and N in silicon nitride," Phys. Rev. B 74, 045203 (2006).

[16] O. Stenzel, S. Wilbrandt, N. Kaiser, M. Vinnichenko, F. Munnik, A. Kolitsch, A. Chuvilin, U. Kaiser, J. Ebert, S. Jakobs, A. Kaless, S. Wüthrich, O. Treichel, B. Wunderlich, M. Bitzer, and M. Grössl, "The correlation between mechanical stress, thermal shift and refractive index in $HfO_2$, $Nb_2O_5$, $Ta_2O_5$ and $SiO_2$ layers and its relation to the layer porosity," Thin Solid Films 517, 6058 (2009).

[17] P. Ambrée, F. Kreller, R. Wolf, and K. Wandel, "Determination of the mechanical stress in plasma enhanced chemical vapor deposited $SiO_2$ and SiN layers," J Vac Sci Technol B Microelectron Nanometer Struct Process Meas Phenom 11, 614 (1993).

[18] A. Fourrier, A. Bosseboeuf, D. B. D. Bouchier, and G. Gautherin, "Annealing effect on mechanical stress in reactive ion-beam sputter-deposited silicon nitride films," Jpn. J. Appl. Phys. 30, 1469 (1991).